%% file: main.tex
\newcommand{\CLASSINPUTtoptextmargin}{0.79in}
\newcommand{\CLASSINPUTbottomtextmargin}{0.99in}
\documentclass[conference]{IEEEtran}
\IEEEoverridecommandlockouts
\usepackage[export]{adjustbox}
\input{sections/preamble}

\def\BibTeX{{\rm B\kern-.05em{\sc i\kern-.025em b}\kern-.08em T\kern-.1667em\lower.7ex\hbox{E}\kern-.125emX}}
\begin{document}
\input{figs/colors}

\title{
Turbo Equalization with Coarse Quantization\\using the Information Bottleneck Method
\vspace{-0.2cm}
}

\author{\IEEEauthorblockN{Philipp Mohr, Jasper Brüggmann, Gerhard Bauch}
\IEEEauthorblockA{\textit{Institute of Communications} \\
\textit{Hamburg University of Technology}\\
Hamburg, Germany \\
\{philipp.mohr; jasper.brueggmann; bauch\}@tuhh.de}
}

\maketitle

\begin{abstract}
This paper proposes a turbo equalizer for intersymbol interference channels (ISI) that uses coarsely quantized messages across all receiver components.
Lookup tables (LUTs) carry out compression operations designed with the information bottleneck method aiming to maximize relevant mutual information.
The turbo setup consists of an equalizer and a decoder that provide extrinsic information to each other over multiple turbo iterations.
We develop simplified LUT structures to incorporate the decoder feedback in the equalizer with significantly reduced complexity.
The proposed receiver is optimized for selected ISI channels.
A conceptual hardware implementation is developed to compare the area efficiency and error correction performance.
A thorough analysis reveals that LUT-based configurations with very coarse quantization can achieve higher area efficiency than conventional equalizers.
Moreover, the proposed turbo setups can outperform the respective non-turbo setups regarding area efficiency and error correction capability.
\end{abstract}

\input{sections/content}
\end{document}

%% file: sections/preamble.tex
\usepackage{cite}
\usepackage{amsmath,amssymb,amsfonts}
\usepackage{graphicx}
\usepackage{textcomp}
\usepackage{xcolor}
\usepackage{pgfplots}
\usepackage{bm}
\usepackage{array}
\usepackage{tikz}
\usepackage{nicefrac}
\usepackage[caption=false]{subfig} %
\usepackage{mathtools} %
\usepackage{comment}    %
\usetikzlibrary{fit}
\usetikzlibrary{calc, shapes.gates.logic.US, shapes.gates.logic.IEC}
\usetikzlibrary{positioning}
\usetikzlibrary{spy,backgrounds}
\usetikzlibrary{shapes.multipart}
\usetikzlibrary{positioning,calc}
\usepgfplotslibrary{fillbetween}
\usetikzlibrary{patterns, shapes.geometric}
\colorlet{mydarkblue}{blue!30!black}
\usepackage{nicefrac} %
\usepackage{multirow}  
\usepackage{placeins} %
\usepackage{soul}   %
\usepackage{algorithm}
\usepackage{algpseudocode}
\usepackage{tabularray}
\usepackage{standalone}
\usepackage{circuitikz}

\def\BibTeX{{\rm B\kern-.05em{\sc i\kern-.025em b}\kern-.08em
    T\kern-.1667em\lower.7ex\hbox{E}\kern-.125emX}}



%% file: figs/colors.tex
\definecolor{intblack}{HTML}{000000}%
\definecolor{intdarkblue}{HTML}{00458A}%
\definecolor{intdarkgreen}{HTML}{197A84}%
\definecolor{intlightgreen}{HTML}{57BDC3}%
\definecolor{intgreen}{HTML}{95BA4F}%
\definecolor{intorange}{HTML}{EC6E00}%
\definecolor{intred}{HTML}{C00000}%
\definecolor{intgreenintense}{HTML}{009933}%
\definecolor{darkblue}{rgb}{0,0.2706,0.541}%
\definecolor{darkgreen}{rgb}{0.098,0.4784,0.5176}%
\definecolor{lightgreen}{rgb}{0.3412,0.7411,0.7647}%
\definecolor{orange}{rgb}{0.9255,0.4313,0}%
\definecolor{darkred}{rgb}{0.7529,0,0}%
\definecolor{black}{rgb}{0,0,0}%

%% file: sections/content.tex
\section{Introduction}
Transmission of digital data streams over channels with memory can cause intersymbol interference (ISI) in the received signal.
Turbo equalization is a well-known technique where an equalizer and a decoder cooperate to mitigate interference with near-capacity error correction performance~\cite{tuchler02turbo}.
However, high implementation complexity arises when using optimal equalization and decoding algorithms~\cite{bahl74opt}.

In \cite{mohr22} a new type of equalization has been proposed that replaces computationally intensive parts of an arithmetic-based forward-backward equalization algorithm with lookup tables (LUTs) designed using the information bottleneck (IB) method.
The IB method is a clustering framework for designing compression operations that maximize mutual information~\cite{lewandowsky2020information}.
The LUTs realize compression operations to reduce the space-complexity of discrete messages exchanged within the IB equalization algorithm.

\begin{figure}[t]
 	\centering
       \includegraphics{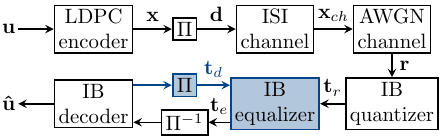}
 	\caption{Iterative equalization and decoding designed using the IB method.}
 \label{fig:setup_proposed}
 \vspace{-0.5cm}
 \end{figure}

This work combines the turbo principle with the IB equalizers as depicted in Fig.\,\ref{fig:setup_proposed}.
By introducing new LUT decompositions, the turbo feedback from the decoder is integrated into the IB equalization algorithm with low complexity.
Compared to the non-turbo setup, significant performance improvements are observed for a magnetic recording and faster-than-Nyquist channel.
Another contribution is the low-level hardware complexity comparison of conventionally quantized equalizers and IB equalizers with turbo and non-turbo configurations.
The computational and memory complexity of all setups is compared using logic gate representations.
It is shown that configurations with very coarse quantization can achieve higher area efficiency than the conventional approaches.

\section{Design of Coarsely Quantized Equalization with Decoder Feedback} \label{sec:design}
A sequence of symbols $\mathbf{d}{=}[.., d_{-1}, d_0, d_1, ..]$ with mapping alphabet $d_k{\in}\mathcal{D}$ is transmitted over an ISI channel with taps $[h_0,h_1,.., h_L]{\in} \mathbb{R}^{L{+}1}$. 
The channel state $s_k$ models the symbols in memory $[d_{k-L},..,d_{k-1}]$.
The receiver observes the sequence $\mathbf{r}{=}[.., r_{-1}, r_0, r_1, ..]$ with $r_k{=}x^{ch}_k{+}n_k$
where $x^{ch}_k{=}\sum_{l{=}0}^{L} h_ld_{k-l}$ and $n_k$ is additive white Gaussian noise (AWGN) with variance $N_0/2$.

Using the BCJR algorithm\cite{bahl74opt}, optimal soft-output equalization for symbol $d_k$ is achieved by computing
\begin{align}
	p(d_k|\mathbf{r})\propto\sum_{\mathclap{s}}\alpha(s|\mathbf{r}_-)\gamma(s,d_k|r_k)\beta(s'|\mathbf{r}_+').
	\label{equ:p_d_given_received}
\end{align}
with the current state $s {\coloneqq} s_k$, the next state $s' {\coloneqq}  s_{k+1}$, the current received symbol $r_k$, all previous received symbols $\boldsymbol{r}_-{\coloneqq} [r_{l}{:}l{<}k]$ and all following received symbols $\boldsymbol{r}_+'{\coloneqq} [r_{l}{:}l{>}k]$.
The branch metric $\gamma(s,d_k|r_k)$ is equal to
\begin{align}
	p(r_k|s,d_k)p(d_k)=\frac{1}{\sqrt{\pi N_0}} \exp\left(\frac{-|r_k-x_k^{ch}|^2}{N_0}\right)p(d_k).
\end{align}
The probability $p(d_k)$ can be considered as extrinsic information from the decoder.
In the first equalizer run this probability corresponds to the prior probabilities.
The forward and backward metrics, $\alpha(s|\mathbf{r}_-)$ and $\beta(s'|\mathbf{r}_+')$ in (\ref{equ:p_d_given_received}), are computed recursively with $\boldsymbol{r}_-'{\coloneqq} [r_{l}{:}l{\le}k]$ and $\boldsymbol{r}_+{\coloneqq} [r_{l}{:}l{\ge}k]$ according to
\begin{align}
	\alpha(s'|\mathbf{r}_{-}')&=\sum_{\mathclap{d_{k-L}}}\gamma(s,d_k|r_k)\alpha(s|\mathbf{r}_{-})\label{equ:forward} \text{ and }\\
	\beta(s|\mathbf{r}_{+})&=\sum_{d_k}\gamma(s,d_k|r_k)\beta(s'|\mathbf{r}_{+}').\label{equ:backward}
\end{align}
\subsection{Forward and Backward Updates with Coarse Quantization}
In \cite{mohr22} it has been proposed to replace the high-resolution vector metrics $\alpha(s|\mathbf{r}_{-})$ and $\beta(s'|\mathbf{r}'_{+})$ with finite alphabet messages $t_\alpha{\in}\mathcal{T}^{w_\alpha}$ and $t'_\beta{\in}\mathcal{T}^{w_\beta}$ using $\mathcal{T}^w{=}\{1,{\ldots},2^{w}\}$ with $w$ bits.
Furthermore, the channel message, extrinsic feedback message from the decoder and message from the equalizer are assumed to be quantized as $t_{r}{\in}\mathcal{T}^{w_r}$, $t_{d}{\in}\mathcal{T}^{w_{d}}$ and $t_{e}{\in}\mathcal{T}^{w_{e}}$.
Hence, the computations in (\ref{equ:forward}) and (\ref{equ:backward}) can be replaced with lookup operations defined by the deterministic conditional probabilities $p(t_{\alpha}'|t_\alpha,t_r, t_{d})$ and $p(t_{\beta}|t_\beta',t_r, t_{d})$.
The factor graph with quantized exchanged messages is shown in Fig.\,\ref{fig:factor_graph_compressed}.
\begin{figure}
	\includegraphics{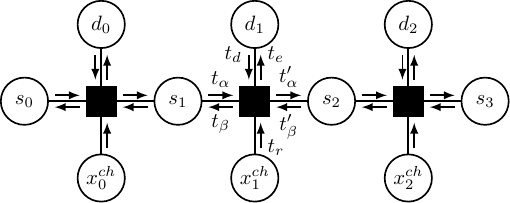}
	\caption{Factor graph with compressed exchanged messages.}
	\vspace{-0.5cm}
	\label{fig:factor_graph_compressed}
\end{figure}
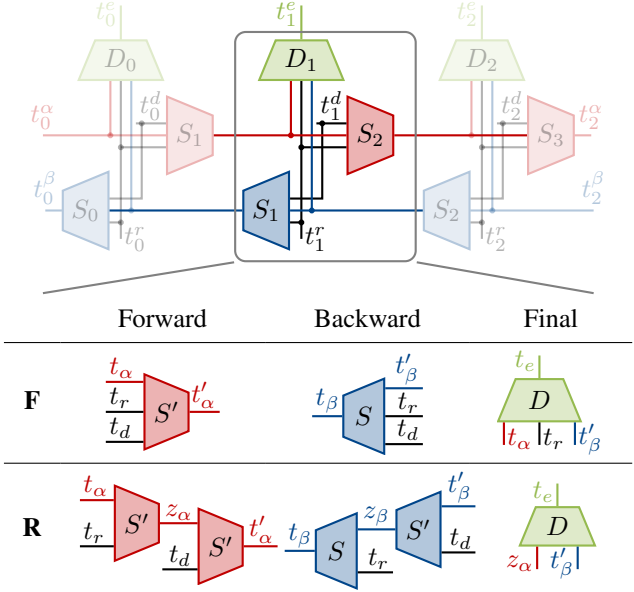
\begin{figure}
	\input{figs/node_structures/overview_src}
	\caption{IB graph and LUT decompositions.}
	\vspace{-0.5cm}
	\label{fig:eq_node_decompositions}
\end{figure}
\begin{figure}[t]
	\centering
	\includegraphics{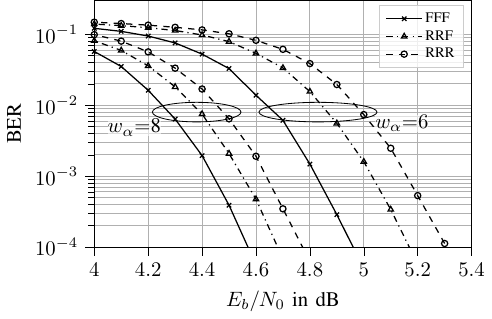}
	\vspace{-0.3cm}
	\caption{%
		Performance of simplified structures for forward, backward and final update. Each update can use a full (F) or reduced (R) LUT structure.}
	\label{fig:ber_decomposition}
	\vspace{-0.3cm}
\end{figure}
All compressed messages $t_\alpha$, $t'_\beta$, $t_r$ and $t_{d}$ can be modeled as realizations of the random variables $T_{\alpha}$, $T'_{\beta}$, $T_{r}$ and $T_{d}$, respectively.
Similarly, the symbol $d_k$, current state $s$ and next state $s'$ are modeled as realizations of the random variables $D$, $S$ and $S'$, respectively.

The forward lookup shall maximize the preserved mutual information between $S'$ and $T_{\alpha}'$ according to 
\begin{align}
    \max_{p(t_{\alpha}'|t_{\alpha}, t_r, t_{{d}})} I(S'; T_{\alpha}').\label{equ:max_mi_forward}
\end{align}
The objective (\ref{equ:max_mi_forward}) can be solved using an IB algorithm where $[T_{\alpha}, T_r, T_{{d}}]$, $S'$ and $T_{\alpha}'$ are classified as the observed variable $Y$, relevant variable $X$ and compressed variable $T$, respectively\cite{mohr22}.
The required input joint distribution $p(x,y)$ for the design is obtained as
\begin{align}
	    p(s',t_{\alpha}, t_r, t_{{d}})=\sum_{\mathclap{d_{k{-}L}}}p(t_r|s, d_k)p(d_k,t_{{d}})p(s, t_{\alpha}).
\end{align}
In Fig.\,\ref{fig:eq_node_decompositions} the factor graph of Fig.\,\ref{fig:factor_graph_compressed} is depicted as an IB graph. The trapezoidal nodes reflect an IB setup where the inputs are the observed variables, the label inside of the node is the relevant variable and the output is the compressed variable~\cite{mohr22}.

The design procedure must keep track of the underlying probability distribution
\begin{align}
    p(s', t_{\alpha}')=\sum_{\mathclap{t_{\alpha}, t_r, t_{{d}}}}p(t_{\alpha}'|t_{\alpha}, t_r, t_{{d}})p(s',t_{\alpha}, t_r, t_{{d}}).
\end{align}
The forward LUTs are designed over multiple recursions. In the first design step, $p(s){=}\nicefrac{1}{{|\mathcal{D}|}^{L}}$ and $p(t_{\alpha}|s){=}\nicefrac{1}{2^{w_\alpha}}$.
After a certain amount of recursions (around 10 to 50), the mutual information $I(S';T_{\alpha}')$ converges.
Then, a post processing step, proposed as the \emph{static design phase} in \cite{mohr22}, finds a single LUT that can be used in all recursions.
 
Similarly, the backward mapping $p(t_{\beta}|t_{\beta}', t_r, t_{{d}})$ is designed using the joint distribution
\begin{align}
	p(s,t_{\beta}', t_r, t_{{d}})=\sum_{d_k}p(t_r|s, d_k)p(d_k,t_{{d}})p(s', t_{\beta}').
\end{align}
The backward output joint distribution yields
\begin{align}
	p(s, t_{\beta})=\sum_{\mathclap{t_{\beta}, t_r, t_{{d}}}}p(t_{\beta}|t_{\beta}', t_r, t_{{d}})p(s,t_{\beta}', t_r, t_{{d}}).
\end{align}

After the static design phase for forward and backward LUTs, a single final mapping $p(t_e|t_{\beta}',t_{r},t_{\alpha})$ is designed using
\begin{align}
	p(d_k, t_\alpha, t_r, t_{\beta}')=\sum_{d_k}p(t_{\beta}'|s')p(s, d_k,t_r)p(t_\alpha|s)
\end{align}
where $p(t_\alpha|s)$ and $p(t_{\beta}'|s')$ are obtained after a certain amount of recursions with the static LUTs. Then, the output joint distribution assumed for the decoder design yields
\begin{align}
	p(d_k, t_e)=\sum_{\mathclap{t_{\beta}',t_{r},t_{\alpha}}}p(t_e|t_{\beta}',t_{r},t_{\alpha})p(d_k, t_\alpha, t_r, t_{\beta}').
\end{align}
\subsection{Decomposition of Three-Input Tables}\label{sec:decomposition}
\begin{table}[t]
	\caption{Number of LUT entries ($w_r{=}5$, $w_d{=}3$).}
	\vspace{-0.2cm}
	\centering %
	\begin{tabular}{c|c|c||c|c}
		&\multicolumn{2}{c||}{Forward/Backward}&\multicolumn{2}{c}{Final}\\
		Variant & 8 bit & 6 bit & 8 bit & 6 bit \\
		\hline 
		Full (F) & \multirow{1}{*}{66k} & \multirow{1}{*}{16k}& 2097k                            &  131k\\
		Reduced (R) &                   \textcolor{black}{10k}        &            \textcolor{black}{2.5k}              & \textcolor{black}{\multirow{1}{*}{66k}}  &  \textcolor{black}{\multirow{1}{*}{4k}}
	\end{tabular}
	\vspace{-0.45cm}
	\label{table:table_size}
\end{table}
Fig.\,\ref{fig:eq_node_decompositions} proposes simplified structures, labeled reduced~(R), that shrink the overall LUT size through concatenation of two-input LUTs. 
The reduction in size causes a performance loss compared to the three-input LUTs, labeled full~(F), investigated in Fig.\,\ref{fig:ber_decomposition}.
The simulation results are obtained for a magnetic recording channel with taps $[.5,.5,-.5,-.5]$, BPSK modulation, a rate $1/2$ regular LDPC code used in \cite{mohr22} and 3 turbo iterations each with 10 decoder iterations.

It can be observed that the decomposition of the forward and backward three-input LUT in variant RRF leads to .1\,dB and .2\,dB loss compared to variant FFF for $w_{\alpha}{=}w_{\beta}{=}8$\,bits and  $w_{\alpha}{=}w_{\beta}{=}6$\,bits, respectively.
As in \cite{mohr22}, the decomposition of the final LUT in variant RRR causes an additional loss of .08\,dB and .13\,dB, respectively.
Thus, accepting moderate performance loss enables major savings in terms of LUT size as shown in Table \ref{table:table_size}. 
The decomposition of forward and backward LUTs reduces the number of LUT entries from $2^{w_\alpha+w_r+w_d}$ to $2^{w_\alpha}(2^{w_r}+2^{w_d})$.
Decomposing the final LUT leads to a reduction from $2^{w_\alpha+w_r+w_\beta}$ to $2^{w_\alpha+w_\beta}$.

\section{Proposed vs. Conventional Equalization} 

Logarithmic probabilities $\bar{p}{\coloneqq}\log p$ are considered for a conventional equalizer.
Then, the forward recursion yields\cite{colavolpe05onmap}
\begin{align}
	\bar{\alpha}(s') = \stackrel{*}{\max_{d_{k{-}L}}} \left(\bar{\alpha}(s)+\frac{1}{N_0}\bar{\gamma}(s,d_k)+ \bar{p}(d_k|t_{d})\right)
	\label{equ:forward_conv_log}
\end{align}
In (\ref{equ:forward_conv_log}), for any two inputs $a,b{\in}\mathbb{R}^2$, we recursively apply\cite{robertson95comp}
\begin{align}
	\stackrel{*}{\max}(a,b)=\max(a,b)+\log(1+e^{-|a-b|}).
	\label{equ:jacobian}
\end{align}
The final update computes the LLR for the decoder as
\begin{align}
	\bar{p}(d_k|\mathbf{r}) = \stackrel{*}{\max_{s}}\left(\bar{\alpha}(s)+\frac{1}{N_0}\bar{\gamma}(s,d_k)+\bar{\beta}(s')\right).
	\label{equ:forward_final_log}
\end{align}
The metric computation $\bar{\gamma}(s,d_k)$ depends on the observation model.
The Forney observation model directly uses the channel observations $r_k$ for computing the branch metric\cite{forney72maximum,colavolpe05onmap}
\begin{align}
	\bar{\gamma}(s,d_k)=-|r_k-x^{ch}_{s,d_k}|^2 \quad \text{ with } x^{ch}_{s,d_k} = \sum\limits_{l{=}0}^{L}h_l d_{k-l}.
	\label{equ:forney}
\end{align}
The Ungerboeck observation model\cite{ungerboeck74adaptive,colavolpe05onmap} applies a prefilter leading to $r'_k{=}\sum_{\kappa}h'_{\kappa}r_{k-\kappa}$ where $h'_l=h_{-l}^*$. Then, 
\begin{align}
	\bar{\gamma}(s,d_k)=2\mathrm{Re}\{d_k\cdot r'_k\} - \gamma_{s,d_k}
	\label{equ:ungerboeck}
\end{align}
with the function $\gamma_{s,d_k} = g_0|d_k|^2+\mathrm{Re}\left\{d^*_k \sum_{l{=1}}^{L}g_l d_{k-l} \right\}$ where $g_k{=}\sum_{i=-L}^{L}h_ih^*_{i-k}$.
Another variant, termed channel shortening, applies a modified prefilter $h'$ leading to a shortened model $g_k$ with smaller memory $L$\cite{rusek12opt}.
However, the approach reduces performance and the modified prefilter involves approximately 30 taps leading to additional complexity.
\subsection{Complexity Reduction of The Conventional Equalizer}\label{sec:conv_complexity_reduction}
The Forney model involve an expensive $|.|^2$-operation in (\ref{equ:forney}) which must be computed for every state transition.
The multiplication implementation in the Ungerboeck model (\ref{equ:ungerboeck}) is very simple with BPSK modulation as $2\mathrm{Re}\{d_k r'_k\}\in \{\pm 2r'_k\}$, also mentioned in~\cite{ungerboeck74adaptive}. Hence, overall the Forney metric demands higher complexity with $3{\cdot}2^{L{+}1}$ multiplications per equalizer run compared to the Ungerboeck model that only uses $L{+}1$ multiplications in the prefilter operation. For a magnetic recording channel with the channel taps $[.5,.5,-.5,-.5]$, the prefilter can avoid multiplications completely.
When neglecting the correction term in (\ref{equ:jacobian}) complexity can be saved.
Then, e.g., the forward recursion simplifies to
\begin{align}
	\tilde{\alpha}(s') \approx \max_{d_{k-L}} \left(\bar{\alpha}(s){+}\bar{\gamma}(s,d_k)\right)+N_0\bar{p}(d_k|t_{d})
	\label{equ:quant_forward}
\end{align}
where $N_0\bar{p}(d_k|t_{d})$ can be implemented with an LUT because the decoder feedback $t_{d}$ is coarsely quantized.
A normalization $\bar{\alpha}(s'){=}\tilde{\alpha}(s'){-}\tilde{\alpha}(s'{=}1)$ limits the numeric range.

The channel message $r_k$ can be uniformly quantized with spacing $\Delta$.
The spacing $\Delta$ is preserved within all equalizer operations.
A clipping operation limits the bit width of every entry in the forward metrics $\bar{\alpha}(s)$ to $\bar{w}_\alpha$. The bit width of the vector metrics is $w_\alpha{=}(2^L{-}1) \bar{w}_\alpha$ as $\bar{\alpha}(s{=}1)$ is always 0.

\subsection{Magnetic Recording Channel}\label{sec:ber_epr_bpsk}
\begin{figure}[t]
	\centering
	\includegraphics{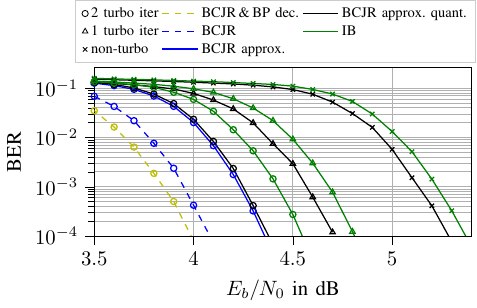}
	\vspace{-0.3cm}
	\caption{EPR4 channel: Conventional vs. IB ($w_\alpha{=}9$).}
	\label{fig:ber_epr_bpsk}
	\vspace{-0.2cm}
\end{figure}

Fig.\,\ref{fig:ber_epr_bpsk} evaluates the error rate performance for a magnetic recording channel with the same setup as in section~\ref{sec:decomposition} using decomposition RRR (cf.~Fig.\,\ref{fig:ber_decomposition}).
The number of turbo iterations $N_{it}$ is configured as 2,1 or 0. Similar decoder complexity is established across all setups by distributing a total budget of 20 decoder iterations as  $(5, 5, 10)$, $(10, 10)$ and $(20)$, respectively. 
If not mentioned otherwise, all setups use LUT-based decoders with 4\,bit for the exchanged messages as in~\cite{mohr22}.
The decoder uses the existing messages in memory from previous decoder runs.

The non-turbo IB equalizer proposed in \cite{mohr22} is outperformed by up to .55 and .8\,dB when using 1 or 2 turbo iterations, respectively.
The IB equalizers are configured with 5\,bits for the channel message and 9\,bits for the state metrics.

A conventional equalizer requires 7\,bits for the channel message and 11\,bits for each state metric entry (in total 77\,bits with 8 states) to maintain similar performance as its non-quantized variant.
Although the number of bits is decreased by an order of magnitude, the IB equalizers suffer only from .03\,dB, .05\,dB and .1\,dB  performance loss compared to the conventional equalizer at 0, 1 and 2 turbo iterations, respectively.
The increasing difference can be explained by the decomposition structure RRR, which is used more frequently according to the number of turbo iterations causing a higher cumulative loss.

The conventional equalizer performance can be improved by .27\,dB using the accurate $\stackrel{*}{\max}$ implementation.
Further gains of .1\,dB in the configuration 'BCJR\&BP dec.' are obtained using a high-resolution belief propagation LDPC decoder instead of a 4-bit LUT decoder.
\subsection{Faster-Than-Nyquist Channel}
\begin{figure}[t]
	\centering
	\includegraphics{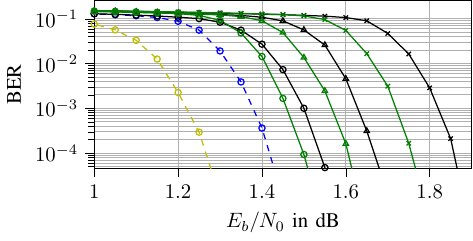}
	\vspace{-0.3cm}
	\caption{FTN channel: Conventional vs. IB ($w_\alpha{=}8$).}
	\label{fig:ber_ftn_bpsk}
	\vspace{-0.2cm}
\end{figure}

Fig.\,\ref{fig:ber_ftn_bpsk} depicts the error correction performance for a channel with ISI resulting from faster-than-Nyquist (FTN) signaling as in~\cite{mohr22}.
A 64800-length DVB-S2 LDPC code with rate $1/2$ is used.
The number of decoder iterations is distributed as $(50)$, $(25,25)$ and $(16,17,17)$ for 0, 1 and 2 turbo iterations, respectively.
As the impulse response is long $\mathbf{h}{=}$[.8907,.4088,-.1919,.0510,-.0040, .0045,-.0076,.0039,-.0014,.0019,-.0020,.0014], all equalizers perform a shortening approach to reduce the equalizer memory.
The conventional equalizers perform channel shortening leading to 4 states (memory 2).
The IB equalizers truncate the model after 3 taps.
The non-turbo IB equalizer \cite{mohr22} is outperformed by .17\,dB and .27\,dB when using 1 or 2 turbo iterations, respectively.
The gains are smaller  than in the previous section~\ref{sec:ber_epr_bpsk} as the ISI is less severe.
All quantized BCJR equalizers with the approximated $\stackrel{*}{\max}$ implementation are outperformed by .2\, .1, and .07\,dB for 0, 1 and 2 turbo iterations.

\section{Hardware Complexity}
\newcommand{\includeminipage}[1]{{\begin{minipage}{\linewidth}\centering\includegraphics{\currfiledir #1} \end{minipage}}}

\begin{figure}[t]
	\subfloat[\centering Division into sub-blocks for increasing parallelism.]{\includegraphics{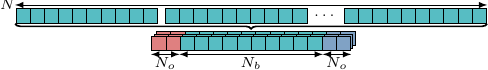}	\label{fig:subblocks}}\vspace{-0.1cm}
	\hfil
	\subfloat[\centering 3 unrolled turbo iterations using an X-shaped structure\cite{weithoffer2018years}.]{\includegraphics{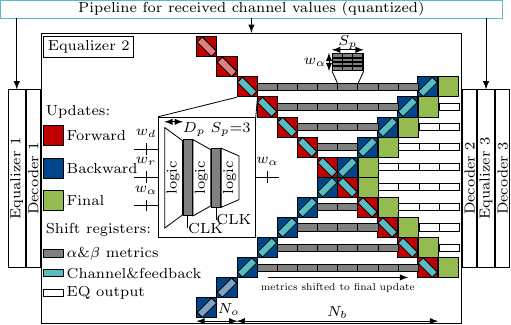}	\label{fig:eq_dec_shift_register}}%
	\caption{Turbo equalizer setup with $N_b{=}10$ and $N_o{=}2$.}
	\vspace{-0.3cm}
\label{fig:forward_conv_adder}
\end{figure}

This section estimates and compares the hardware complexity of the LUT-based and arithmetic-based equalizers.
The complexity is typically expressed in terms of area efficiency, i.e., the ratio of throughput to chip area.
The forward and backward recursions can cause high delays when equalizing $N$ symbols sequentially.
To reduce the delay and to improve throughput, $P$ sub-blocks with length $N_b$ are equalized in parallel as shown in Fig.\,\ref{fig:subblocks}. 
Each sub-block involves an initialization phase using $N_o$ overlapping symbols of the adjacent sub-blocks.
It can be observed in \cite{mohr22} that an overlap length of $N_o{=}10$ is sufficient to avoid performance loss for a magnetic recording channel.
A conceptual hardware implementation for 3 unrolled turbo iterations is depicted in Fig.\,\ref{fig:eq_dec_shift_register}.
The forward and backward recursions can be carried out using an X-shaped structure~\cite{weithoffer2018years}. 
We divide the logic gates that carry out the update operations into a number of levels $N_L$.
The logic depth per pipeline stage $D_p$ defines the number of logic gate levels between a pair of consecutive pipeline stages.
A lower value for the logic depth $D_p$ allows higher clock frequency and better logic utilization at the cost of additional shift registers. %
We use a small value $D_p{=}8$ throughout the analysis to ensure high logic utilization. 
In this way, it is assumed that all implementations use the same clock frequency. 
This means that the area efficiency only depends on the chip area.

The chip area is determined by the number of transistors used to realize the update logic, the pipeline stages and the shift registers for storing the state metrics. In this work, interconnects are neglected to avoid extensive hardware synthesis and dependence on a specific hardware platform.
Table \ref{table:complexity_components} lists the assumed transistor count of logic operations.

The overall transistor count for $N_{it}$ turbo iterations is 
\begin{align}
	\xi_{eq}^{(N_b,N_o)}=\sum_{i=0}^{N_{it}}\left(\xi_{update}^{(i,N_b)}+\xi_{memory}^{(i,N_b, N_o)}\right)
\end{align}
where the update logic and memory logic are normalized by the number of equalizer outputs $N_b$ per clock cycle as
\begin{align}
	\xi_{update}^{(i,N_b, N_o)}&{=}\frac{1}{N_b}\left((N_b{+}N_o)(\xi^{(i)}_{\alpha}{+}\xi_{\beta}^{(i)}){+}N_b\xi_{e}^{(i)}\right) \text{ and }\\
	\xi_{memory}^{(i,N_b)}&{=}\frac{1}{N_b}(S_{p}(w_{\alpha}^{(i)}{+}w_{\beta}^{(i)}){+}S_{p,e}w_e)\xi_{DFF}\sum_{k=1}^{\mathclap{N_b/2-1}}2k\,.\end{align}
The transistor count for forward, backward and final update (a single square box in Fig.\,\ref{fig:eq_dec_shift_register}) is $\xi^i_{\alpha}$, $\xi_{\beta}^i$ and $\xi_{e}^i$, respectively, also comprising chip area for the pipeline stages.
The memory complexity $\xi_{memory}^{(i,N_b)}$ increases linearly with the number of pipeline stages~$S_{p}{=}\lceil N_L/D_p \rceil$ as each pipelined sub-block requires dedicated shift registers for the metrics.
The shift registers consist of multiple D-Flipflops (each 4 NAND, 1 NOT, $\xi_{DFF}{=}18$).
The choice of $N_b$ is a compromise between update and memory complexity. 
We remark that the latency changes with $N_b$, but is not evaluated further since the area efficiency is our main focus. 
Thus, in each setup individual optimization of the sub-block length is done according to
\begin{align}
	\min_{N_b}\xi_{eq}^{(N_b,N_o{=}10)}.\label{equ:lenoptimization}
\end{align}

\subsection{Proposed Hardware with Lookup Operations}
\newcommand{\gateimg}[1]{\begin{tikzpicture}[label distance=2mm]\node[#1 gate US, draw, rotate=-90,scale=0.7]{};\end{tikzpicture}}
\begin{table}[t]
	\caption{Required CMOS transistors per logic gate \cite{gajda2007reducing}.} 
	\vspace{-0.2cm}
	\centering
	\begin{tabular}{c|c|c|c|c}
		AND \gateimg{and}& OR\gateimg{or} & NOT \gateimg{not} & NAND \gateimg{nand} & XOR\gateimg{xor} \\ \hline
		6 & 6 & 2 & 4 & 10 \\
	\end{tabular}
	\vspace{-0.4cm}
	\label{table:complexity_components}
\end{table}
\begin{figure}[t]
	\centering
	\subfloat[\centering Decomposition into two LUTs.]{\begin{minipage}{\linewidth}\centering\includegraphics{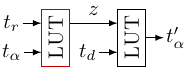}	\label{fig:forward_proposed_decomposition}\end{minipage}}\vspace{-0.3cm}
	\hfil
	\subfloat[\centering LUT with multiplexer implementation.]{\includegraphics{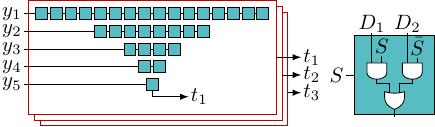}	\label{fig:forward_proposed_lut_mux}}\vspace{-0.3cm}
	\hfil
	\subfloat[\centering LUT with two-level and multi-level implementation.]{\begin{minipage}{\linewidth}\centering\resizebox{0.7\linewidth}{!}{\includegraphics{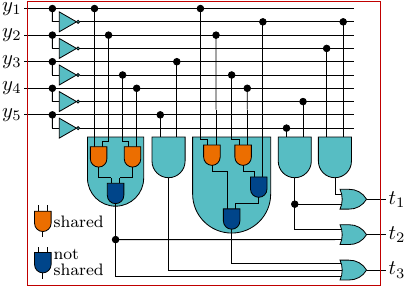}}	\label{fig:forward_proposed_lut_dnf}\end{minipage}}%
	\caption{LUT-based equalizer operations ($w_\alpha=3,w_r=2$).}
	\vspace{-0.5cm}
	\label{fig:forward_proposed}
\end{figure}

\begin{figure}[t]
	\subfloat[\centering Conventional hardware structure of the forward update with 4 states.]{
		\includegraphics[width=0.5\linewidth,valign=c]{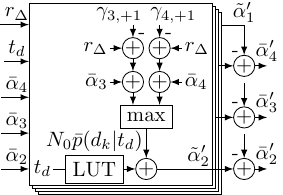}
		\label{fig:hardware_structures_conv}}%
	\hfil
		\subfloat[\centering Ladner-Fischer adder \cite{koren2001computer}.]{
			\includegraphics[width=0.4\linewidth,valign=c]{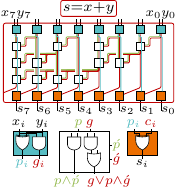}
				\label{fig:hardware_structures_conv_adder}}%
	\hfil
	\caption{Arithmetic-based equalizer implementation.}
	\vspace{-0.5cm}
	\label{fig:forward_conv}
\end{figure}

In the LUT-based equalizer the forward recursion performs a lookup operation with the forward message $t_\alpha$ from the previous recursion step, the quantized channel message $t_r$ and the feedback message $t_{d}$ from the decoder.
As discussed in section~\ref{sec:decomposition}, the LUT size is significantly reduced with a decomposition into two LUTs (cf.~Fig.\,\ref{fig:forward_proposed_decomposition}). 
\subsubsection{Implementation with Multiplexers}
In Fig.\,\ref{fig:forward_proposed_lut_mux}, the LUT is implemented as a selection network, generating a $w_{\alpha}$-bit output $t$ for the input $y$ using  a tree structure. 
The bit values of $y$ are denoted by $y_{1}$ to $y_{w}$.
The LUT entries $\operatorname{LUT}_1$ to $\operatorname{LUT}_{2^{w}}$ initialize the selection process.
In the first stage, $y_1$ causes each 2:1 multiplexer to forward one of its two inputs from the memory cells.
In the next stages, $y_2$ to $y_w$ further proceed the selection until the output $t$ is generated. Hence, $w_\alpha{+}w_r$ multiplexer stages are required.

The second LUT operates similarly using the $w_{\alpha}$-bit intermediate message $t$ and the $w_{d}$-bit feedback message $t_{d}$ from the decoder, requiring another $w_\alpha{+}w_{d}$ multiplexer stages. However typically, $w_{d}{<}w_{r}$. Hence, the first LUT requires most of the complexity.
\subsubsection{Implementation with Two-Level Logic}
The LUT behavior can be represented using logic expressions.
In this paper we consider the disjunctive normal form (DNF) generated by constructing a truth table defining a bit-wise input-output relationship.
We make use of a logic optimization tool~\cite{drake2015pyeda} that implements the Espresso algorithm~\cite{brayton1984logic} to approach a minimal expression of the DNF.
Fig.\,\ref{fig:forward_proposed_lut_dnf} depicts an exemplary implementation of the minimized DNF using logic AND as well as OR gates.
\subsubsection{Implementation with Shared Multi-Level Logic}
The multi-input AND gates in Fig.\,\ref{fig:forward_proposed_lut_dnf}, termed nodes, can be realized with several two-input AND gates.
Among the nodes, some of the internal operations, indicated by the orange color in Fig.\,\ref{fig:forward_proposed_lut_dnf}, have the same input signal.
To improve efficiency, all those operations can be replaced with a single operation whose output signal is utilized by multiple nodes.
We developed an effective algorithm that maximizes utilization of the single operations.
In each iteration step, the algorithm selects the AND combination of two signals that maximizes the number of subsequent operations using the combined signal.
The algorithm terminates when all inputs are combined.

\subsubsection{Exploiting channel symmetries}
As proposed in \cite{mohr22}, channel symmetries can be used to enforce a symmetric design of the LUT.
Then, the LUT can be implemented as
\begin{align}
	t=\operatorname{LUT}(y)=\begin{dcases}
	\operatorname{LUT}'(y_{2:w}) & y_1=0\\
	\neg\operatorname{LUT}'(\neg y_{2:w}) & y_1=1
	\label{equ:symlookup}
\end{dcases}
\end{align}
where $y_{2:w}$ denotes the bits $y_2$ to $y_w$ and $\neg$ is a bit inversion. 
Halving the LUT size saves $2^{w{-}1}$ multiplexers in Fig.\,\ref{fig:forward_proposed_lut_mux} and approximately halves the gate count in~Fig.\,\ref{fig:forward_proposed_lut_dnf}.
The conversion (\ref{equ:symlookup}) requires only $w{-}1$ multiplexers at input and output.

\subsection{Comparison to Conventional Equalizer Implementations}
		\begin{figure}[t]
			\centering
			\includegraphics{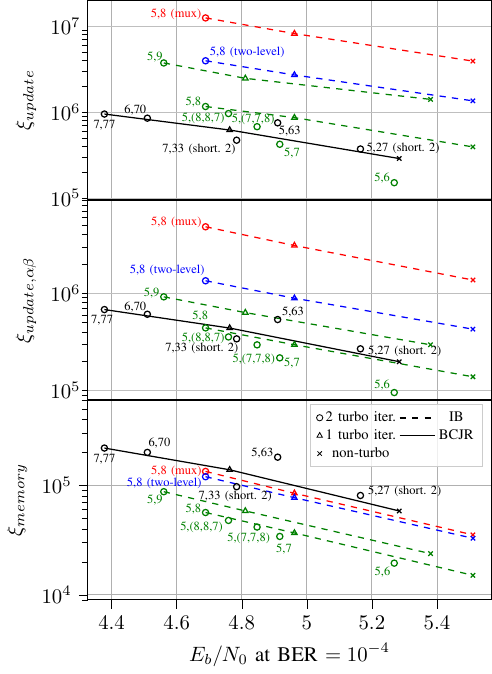}    
			\vspace{-0.3cm}
			\caption{Complexity (transistor count per equalized symbol) vs. performance.}
			\label{fig:num_transistors}
			\vspace{-0.5cm}
		\end{figure}
Fig.\,\ref{fig:hardware_structures_conv} depicts the forward update that implements the behavior of an arithmetic-based log-domain equalizer as in (\ref{equ:forward_conv_log}) and (\ref{equ:forward_final_log}) with simplifications from section \ref{sec:conv_complexity_reduction}. Most of the complexity lies within the adder structure shown in Fig.\,\ref{fig:hardware_structures_conv_adder}, also realizing major part of the maximum operation.

For comparing complexity and performance, we assume a magnetic recording channel setup (cf. section \ref{sec:ber_epr_bpsk}).
In Fig.\,\ref{fig:num_transistors} the number of transistors for various equalizer parts is depicted versus $E_b/N_0$ at a target bit error rate of $10^{-4}$. The configurations are labeled with the number of bits $(w_r,w_{\alpha/\beta})$ used for the channel message and forward/backward metrics. The IB setups (5,(8,8,7)) and (5,(7,7,8)) use different bit widths for the individual turbo iterations and demonstrate another option for trading complexity and performance.

The complexity for the update $\xi_{update}$ reveals that only the IB equalizer with shared logic gates (cf. Fig.\,\ref{fig:forward_proposed_lut_dnf}) can achieve similar or even lower implementation complexity than the arithmetic-based equalizers.
For example, the (5,6) setup with 3 turbo iterations requires half the complexity of the conventional (7,77) setup without turbo iterations at similar performance.

The final LUT is significantly larger than the forward/backward LUT according to table \ref{table:table_size}. Hence, considering only the forward/backward update complexity with $\xi_{update,\alpha\beta}$ even the higher bit width (5,8) setup can achieve lower complexity than the conventional forward/backward update.
Thus, a hybrid design with forward/backward LUTs and final arithmetic update would be a viable option.

Finally, the memory complexity $\xi_{memory}$ shows that all IB equalizers require significantly less memory logic as the forward/backward metrics are smaller by a factor of 3~to~13.
The memory complexity varies among the configurations as it depends on the sub-block length $N_b$ which is optimized in~(\ref{equ:lenoptimization}).
Reducing the bit width of the conventional equalizers causes severe performance degradation for (6,70) or (5,63) while saving only small amounts of complexity.
Instead using channel shortening (7,33,(short. 2)) can offer a better performance-complexity trade-off (neglecting the prefilter in Fig.\,\ref{fig:num_transistors}). %
However, comparing the range of numbers for $\xi_{memory}$ with $\xi_{update}$ clearly shows that the area-efficiency of the forward-backward algorithm is update-bound and not memory-bound.
Thus, the major memory savings with the IB equalizer have only small impact on the overall complexity.

\section{Conclusions}
This paper presented turbo equalizers utilizing two-input LUTs as compression updates to generate coarsely quantized metrics. 
The information bottleneck method was employed for the LUT design.
Significant performance improvements of up to $.8$\,dB were achieved with the turbo over the non-turbo setups. 
The LUT equalizers reduced the memory requirements for buffering metrics by up to an order of magnitude compared to conventional arithmetic-based equalizers.
However, the overall complexity, including the logic for update operations, was only lower under very coarse quantization.

\bibliographystyle{IEEEtran}
\bibliography{literature}

%% file: figs/node_structures/overview_src.tex
\newcommand{\includeminipage}[2]{{\begin{minipage}{#2\linewidth}\centering\includegraphics{\currfiledir #1} \end{minipage}}}
\begin{minipage}{0.9\linewidth}
\begin{tblr}{
      width=\linewidth,
       colspec = { X[1,c,m]  X[7,c,m]  X[7,c,m] X[4,c,m]}
    }
    \SetCell[c=4]{c,m} \includeminipage{equalizer}{1.0} \\
       & Forward & Backward & Final \\ \hline
     \bf{F} & \includeminipage{forward/turbo_A}{1.0} & \includeminipage{backward/turbo_A}{1.0} & \includeminipage{final/full}{1.0} \\ \hline
     \bf{R} & \includeminipage{forward/turbo_C}{1.0} & \includeminipage{backward/turbo_C}{1.0} & \includeminipage{final/reduced_turbo_C}{1.0}
\end{tblr}
\end{minipage}